\documentclass[aps,prb,reprint,superscriptaddress
]{revtex4-2}
\usepackage{amsmath}
\usepackage{amssymb}
\usepackage[utf8]{inputenc}
\usepackage[T1]{fontenc}
\usepackage{mathptmx}
\usepackage{etoolbox}
\usepackage{amsthm}
\usepackage{float}
\usepackage{hyperref}
\usepackage{physics}
\usepackage{subfigure}
\usepackage{svg}
\usepackage{bm}
\usepackage{verbatim}
\usepackage{wrapfig}
\usepackage[space]{grffile}
\usepackage{mathtools}  
\usepackage{xfrac}
\usepackage{graphicx}
\draft 
\usepackage{times}
\newcommand{\angstrom}{\mbox{\normalfont\AA}}
\begin{document}

\preprint{APS/123-QED}

\title{Reconstructing the wavefunction of magnetic topological insulators MnBi$_2$Te$_4$ and MnBi$_4$Te$_7$ using spin-resolved photoemission}%

\author{Xue Han}
\affiliation{Stanford Institute for Materials and Energy Sciences, SLAC National Accelerator Laboratory, Menlo Park, California 94025, USA}
\affiliation{Geballe Laboratory for Advanced Materials, Department of Physics and Applied Physics, Stanford University, Stanford, California 94305, USA}

\author{Jason Qu}
\affiliation{Stanford Institute for Materials and Energy Sciences, SLAC National Accelerator Laboratory, Menlo Park, California 94025, USA}
\affiliation{Geballe Laboratory for Advanced Materials, Department of Physics and Applied Physics, Stanford University, Stanford, California 94305, USA}

\author{Hengxin Tan}
\affiliation{Department of Condensed Matter Physics, Weizmann Institute of Science, Rehovot 7610001, Israel}

\author{Zicheng Tao}
\affiliation{School of Physical Science and Technology, ShanghaiTech University, Shanghai 201210, China}

\author{Noah M. Meyer}
\affiliation{Stanford Institute for Materials and Energy Sciences, SLAC National Accelerator Laboratory, Menlo Park, California 94025, USA}
\affiliation{Geballe Laboratory for Advanced Materials, Department of Physics and Applied Physics, Stanford University, Stanford, California 94305, USA}

\author{Patrick S. Kirchmann}
\affiliation{Stanford Institute for Materials and Energy Sciences, SLAC National Accelerator Laboratory, Menlo Park, California 94025, USA}

\author{Yanfeng Guo}
\affiliation{School of Physical Science and Technology, ShanghaiTech University, Shanghai 201210, China}
\affiliation{ShanghaiTech Laboratory for Topological Physics, Shanghai 201210, China}

\author{Binghai Yan}
\affiliation{Department of Condensed Matter Physics, Weizmann Institute of Science, Rehovot 7610001, Israel}

\author{Zhi-Xun Shen}
\email[]{zxshen@stanford.edu}
\affiliation{Stanford Institute for Materials and Energy Sciences, SLAC National Accelerator Laboratory, Menlo Park, California 94025, USA}
\affiliation{Geballe Laboratory for Advanced Materials, Department of Physics and Applied Physics, Stanford University, Stanford, California 94305, USA}

\author{Jonathan A. Sobota}
\email[]{sobota@slac.stanford.edu}
\affiliation{Stanford Institute for Materials and Energy Sciences, SLAC National Accelerator Laboratory, Menlo Park, California 94025, USA}

\date{\today}

\begin{abstract}
Despite their importance for exotic quantum effects, the surface electronic structure of magnetic topological insulators MnBi$_2$Te$_4$ and MnBi$_4$Te$_7$ remains poorly understood. Using high-efficiency spin- and angle-resolved photoemission spectroscopy, we directly image the spin-polarization and orbital character of the surface states in both compounds and map our observations onto a model wavefunction to describe the complex spin-orbital texture, which solidifies our understanding of the surface band structure by establishing the single-band nature of the most prominent states. Most importantly, our analysis reveals a new mechanism for reducing the magnetic gap of the topological surface states based on the orbital composition of the wavefunction. 
\end{abstract}

\maketitle

\begin{figure}[!t]
    \includegraphics{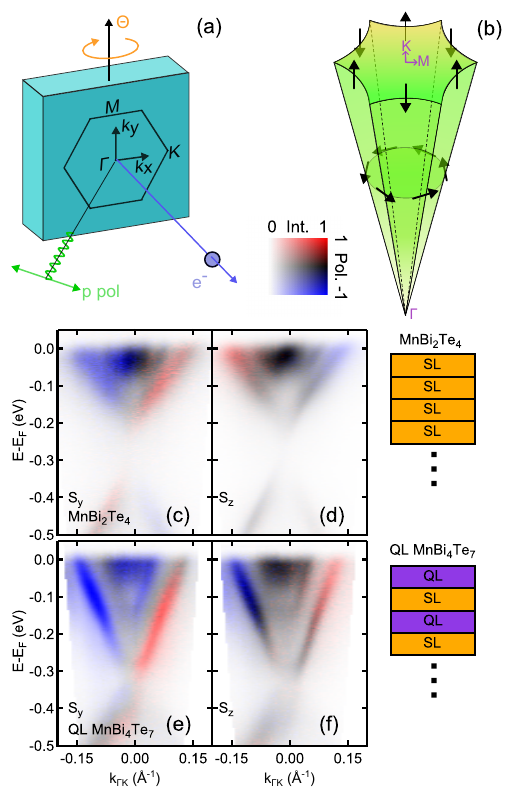}
    \caption{(a) A schematic of the experimental geometry for spin-resolved ARPES measurements. Momentum space is mapped along $k_x$ ($\Gamma-K$) by rotating the sample about the $\theta$-axis. Data in this figure is taken with $p$-polarized light. (b) A simplified illustration of the surface state spin texture in materials with $C_{3v}$ symmetry. Near $\Gamma$ the surface state hosts in-plane helical spin texture, which becomes out-of-plane approaching lower binding energy. In the following, spin-resolved ARPES data are plotted with the 2D colorscale shown, where red-blue denotes the spin-polarization and white-black denotes the photoemission intensity. (c) In-plane $S_y$ and (d) out-of-plane $S_z$ measurement for MnBi$_2$Te$_4$, which consists of septuple layers (SL) stackings. (e,f) Same for quintuple layer (QL) terminated  MnBi$_4$Te$_7$, which consists of alternative stacking of QL and SL. 
    }
    \label{fig:1}
\end{figure}

\begin{figure*}
    \includegraphics{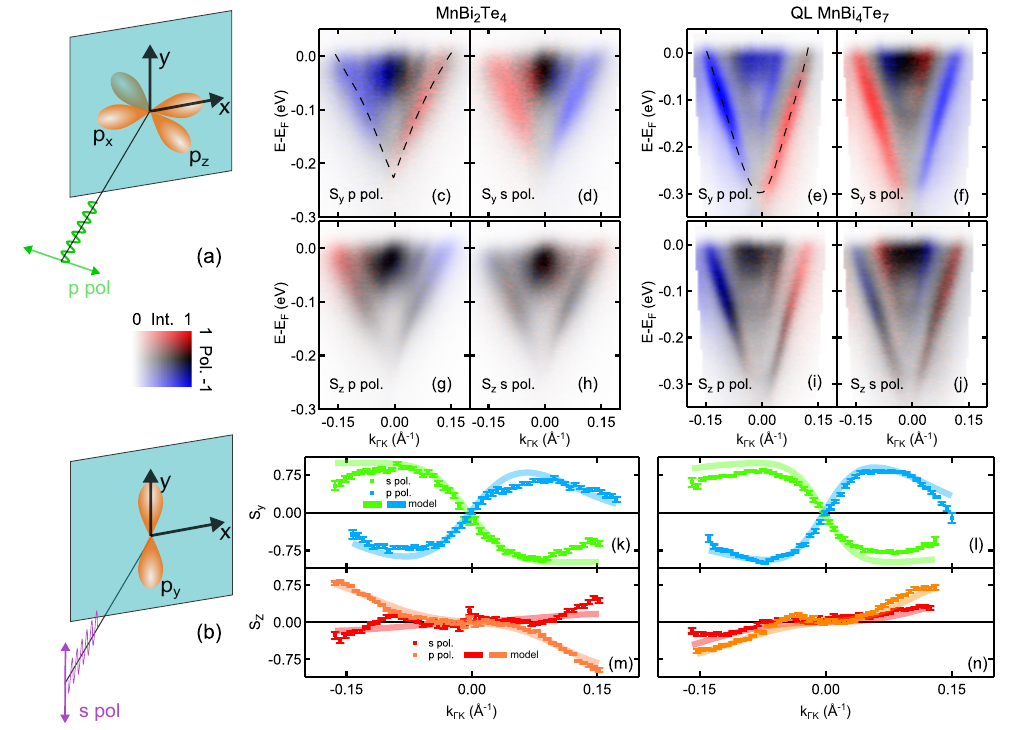}
    \caption{(a) Measurement schematic with  $p$-polarized light, which mainly couples to $p_x$ and $p_z$ orbitals. (b) Similarly, $s$-polarized light mainly couples to $p_y$ orbitals. 
    (c,d) $S_y$ of MnBi$_2$Te$_4$ measured with $p$- and $s$-pol, respectively. (e,f)  Same but for QL MnBi$_4$Te$_7$. (g,h) $S_z$ of MnBi$_2$Te$_4$ measured with $p$- and $s$-pol, respectively. (i,j) Same but for QL MnBi$_4$Te$_7$. 
    (k) $S_y$ extracted from the MnBi$_2$Te$_4$ surface state measured with both light polarizations (markers) overlapped with the model (lines). (l) Same but for QL MnBi$_4$Te$_7$. (m) $S_z$ extracted from the MnBi$_2$Te$_4$ surface state measured with both light polarizations, overlapped with the model. (n) Same but for QL MnBi$_4$Te$_7$. We note that asymmetries in the $p$-pol data with respect to $k=0$ are attributed to the measurement geometry, which is included in the modeling [See Appendix~\ref{sec:spintexture_formalism}].}
    \label{fig:2}
\end{figure*}

\begin{figure*}
    \includegraphics{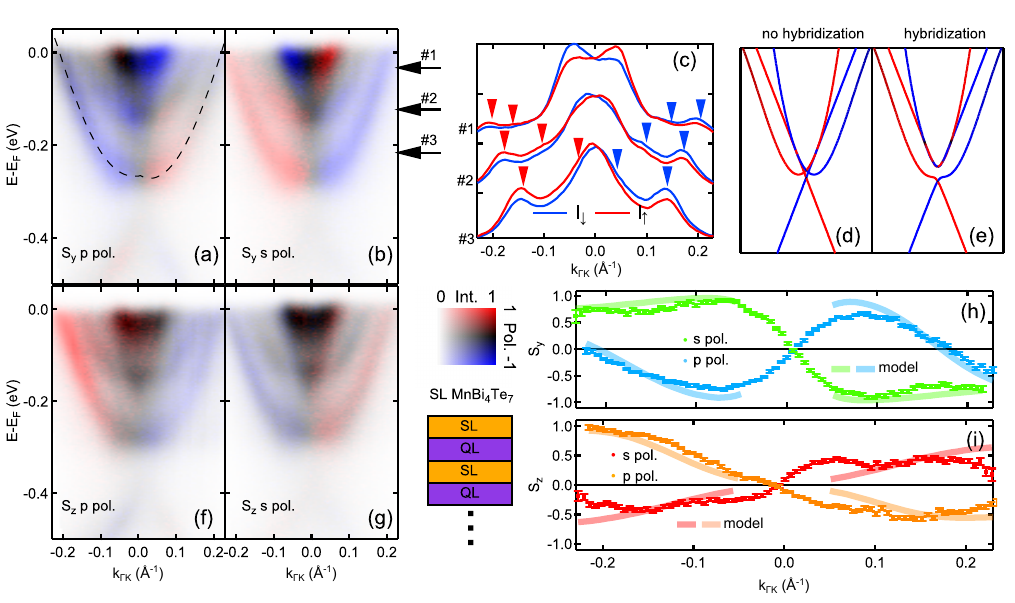}
    \caption{Spin-resolved ARPES measurements of SL MnBi$_4$Te$_7$. (a,b) $S_y$ data with $p$- and $s$-pol light. (c) The corresponding momentum distribution curves (MDCs) of spin-up $I_\uparrow$ ($+y$-direction) and spin-down $I_\downarrow$ ($-y$-direction) intensities. Each MDC is integrated within an energy window of 33~meV.  Each MDC pair is normalized by the maximum intensity for that pair to facilitate a comparison between MDCs at different energies. The markers denote peak positions of both bands, with blue-red coloring indicating the overall polarization of each peak. (d) Cartoon model of a pair of Rashba-split states and Dirac TSS without band hybridization, where color denotes  $S_y$. (e) The same, with hybridization. (f,g) $S_z$ images with $p$- and $s$- pol. (h) $S_y$ extracted along the black dashed line in (a), with $43$~meV integration window for both light polarizations (markers) overlapped with the model (lines). The data within $k=\pm0.05\angstrom^{-1}$ is excluded from the comparison due to the number of overlapping bands in this region. (i) Same but for $S_z$.}
    \label{fig:4}
\end{figure*}

\section{Introduction}
MnBi$_{2n}$Te$_{3n+1}$ (MBT) compounds have been intensively studied in recent years as they constitute the first intrinsic bulk magnetic topological insulator family that has been synthesized \cite{Rienks2019,Otrokov2019,Aliev2019,Klimovskikh2020}. When time-reversal symmetry (TRS) is spontaneously broken, MBT has been shown to host exotic quantum phenomena such as the quantum anomalous Hall effect \cite{Deng2020} and second and third-order nonlinear Hall effects \cite{Wang2023,Gao2023, Li}, 
which result from nontrivial quantum geometry. Despite these consequences of broken TRS, angle-resolved photoemission spectroscopy (ARPES) studies predominantly show a gapless Dirac cone insensitive to the magnetic phase transition \cite{LiHang2019,Hao2019,Chen2019,Hu2020,Nevola2020,Swatek2020, Vidal2021,Wu2020}, in contradiction to theory calculations \cite{Li2019,jiaheng2019, 147Vidal2019,Wu2019,Zhang2019,Shikin2022}.  

In conventional topological insulators (TI) such as Bi$_2$Se$_3$ and Bi$_2$Te$_3$, the measured band structure consists of well-defined bulk and surface states, and therefore the interpretation of even the earliest ARPES data was decisive \cite{Xia2009,Chen2009}. In contrast, ARPES data on MBT exhibits a multitude of bands of uncertain bulk or surface character without clear correspondences in electronic structure calculations, thus precluding a straightforward assignment of the observed bands \cite{Chen2019, Hu2020}. Considering the important controversy surrounding its magnetic behavior, it is critical to first establish a foundational understanding of the electronic structure. 
To provide information beyond the dispersions measured by high-resolution ARPES, we shall demonstrate that probing the structure of the wavefunction itself, including the entangled spin and orbital textures, can be used to help understand the Dirac cone's insensitivity to the magnetic phase transition in MBT. 

 In this work, we perform spin-resolved ARPES measurements on MnBi$_2$Te$_4$ and MnBi$_4$Te$_7$ to systematically track the spin-polarization and orbital character as a function of momentum $k$. By mapping these experimental observations onto a model for the wavefunction and its governing Hamiltonian, we show that the most prominent bands are well-described by a single-band picture dominated by $p$-orbitals and free from hybridization with neighboring bands, reconciling disparate interpretations and solidifying band assignments in the MBT family \cite {Wu2020,Yan2021,Vidal2021, Liang2022,Lee2023}. 
 Our methodology demonstrates that spin- and orbital-resolved ARPES describes the electronic states microscopically and predicts meaningful experimental observables.
Most importantly, we quantitatively show in MnBi$_2$Te$_4$ that a non-negligible contribution of in-plane $p$-orbitals has an antagonistic effect on the surface state gap, thus providing critical insight into the issue of the gapless Dirac cone in the MBT family. 

\section{Spin-orbital texture of topological surface states}

\subsection{Experiment}
The measurement geometry of our spin-resolved ARPES setup is shown in Fig.~\ref{fig:1}(a). We measure along $k_x$ by rotating the sample about the $\theta$ axis, and our high-efficiency spin detector resolves the tangential in-plane spin polarization $S_y$ and out-of-plane spin polarization $S_z$, as described in Appendix~\ref{sec:photomethods} and elsewhere \cite{Qu2023,Han2023}. Photoemission is performed using 6~eV photons with either $s$ or $p$ linear polarization, which we denote as $s$-pol or $p$-pol, respectively.

To establish the baseline electronic structure free from magnetism, the measurements in this work are performed at $80$~K, which is above the N\'{e}el temperatures of MnBi$_2$Te$_4$ ($T_N=24$~K) and MnBi$_4$Te$_7$ ($T_N=13.5$~K) (see Appendix~\ref{sec:synthesis} for synthesis methods). When TRS is preserved, MBT shares the same $C_{3v}$ rotational symmetry as the Bi$_2$Se$_3$ family \cite{Hao2019,Naselli2022}. MnBi$_2$Te$_4$ consists of stacked septuple layers (SL), while MnBi$_4$Te$_7$ consists of alternative stacking of SLs and Bi$_2$Te$_3$ quintuple layers (QL) and therefore exhibits two different surface terminations \cite{Hu2020,Xu2020,Wu2020,HuCW2020,147Vidal2019,Wu2019}. 

Fig.~\ref{fig:1}(b) illustrates the established dispersion and spin texture of a topological surface state (TSS) obeying $C_{3v}$ symmetry. Near $\Gamma$, the energy contour is circular and exhibits helical in-plane spin texture. At higher momenta the energy contour becomes hexagonally warped and develops finite out-of-plane spin polarization along $\Gamma-K$, as has been experimentally observed in Bi$_2$Te$_3$ \cite{Souma2011,Nomura2014,Bentmann2021}. The effective Hamiltonian is given by the third order $k\cdot p$ theory \cite{Fu2009}:
\begin{equation}
\begin{split}
H(k)=&\alpha(k_x\sigma_y-k_y\sigma_x)+\gamma(k_+^3+k_-^3)\sigma_z
\end{split}
\label{eq:kp}
\end{equation}
where $\sigma_x$, $\sigma_y$ and $\sigma_z$ are spin Pauli matrices, and $k_\pm=k_x\pm ik_y$. $\alpha$ defines the in-plane spin-momentum locking, while $\gamma$ introduces the warping and out-of-plane spin. 

Next, we show the in-plane ($S_y$) and out-of-plane ($S_z$) spin-resolved band structures of MnBi$_2$Te$_4$ and QL-terminated MnBi$_4$Te$_7$ in Fig.~\ref{fig:1}. The TSS of both materials exhibits the expected spin-momentum locking, with helical in-plane texture down to the Dirac point as reported previously \cite{Rienks2019,Vidal2019, Wu2020,Shikin2020, Klimovskikh2020,Vidal2021}. In addition, for both surfaces we also observe out-of-plane spin developing at higher $k$, which is well described by the $\gamma$ term in the effective Hamiltonian [Eq.~\ref{eq:kp}]. We note that the $S_z$ polarization direction depends on the sample orientation  (see Fig.~\ref{fig:all124} of Appendix~\ref{sec:Sz}). Additional measurements of the in-plane spin along $\Gamma-M$ confirm the helical spin texture of the TSS, which are shown in Fig.~\ref{fig:GMIP} of Appendix~\ref{sec:GM} . 

While Fig.~\ref{fig:1} suggests that the cartoon in (b) provides a complete description of the TSS, we next show that a richer picture emerges when the orbital degrees of freedom are considered. Based on optical selection rules governing photoemission matrix elements \cite{Zhu2014,Ryoo2016}, measurements using $p$ and $s$ polarized light directly project different orbital components of the wavefunction. Specifically for $p$-orbitals, $p$-pol couples to $p_x$ and $p_z$ orbitals (Fig.~\ref{fig:2}(a)), and $s$-pol selectively photoemits $p_y$ orbitals (Fig.~\ref{fig:2}(b)).

Whereas Fig.~\ref{fig:1} was measured using only $p$-pol, the corresponding measurements using both $p$- and $s$-pol are shown in Fig.~\ref{fig:2}(c-j). To facilitate a quantitative discussion, we also extract the spin-polarizations integrated within a $43$~meV energy window traced along the band dispersions (dashed lines in Fig.~\ref{fig:2}(c,e)), which are plotted using markers in Fig.~\ref{fig:2}(k-n). We highlight several observations: 
(1) $S_y$ behaves qualitatively the same in both compounds, with an overall reversal between $p$- and $s$-pol [Fig.~\ref{fig:2}(k,l)]. 
(2) Depending on the material, $S_z$ may or may not reverse with light polarization [Fig.~\ref{fig:2}(m,n)].
(3) The absolute values of $\lvert S_z \rvert$ measured with $p$- and $s$-pol can be quite different; most strikingly, $\lvert S_z \rvert$ nearly vanishes for $s$-pol measurements on MnBi$_2$Te$_4$, as shown in Fig.~\ref{fig:2}(h) and (m). Such a comparison of $S_z$ measured by $p$- and $s$-pol has never been reported in previous spin-resolved ARPES measurements of conventional TIs \cite{Jozwiak2013,Xie2014,Zhu2014,Bentmann2021}. Similarly, previous spin-resolved ARPES measurements on the MBT family have established spin-momentum locking, but did not reveal this complex orbital-dependent behavior \cite{Rienks2019,Vidal2019, Wu2020,Shikin2020, Klimovskikh2020,Vidal2021}. Such phenomena pose a challenge to our understanding of the wavefunction, therefore, it is necessary to develop a more comprehensive model to describe our observations quantitatively. 

\subsection{Wavefunction model}

We use the effective Hamiltonian shown in Eq.~\ref{eq:kp}. To capture the orbital degrees-of-freedom coupling to spin, we construct the wavefunction basis following the methodology of Ref.~\cite{Zhang2013}, which preserves total angular momentum $J_z=\pm 1/2$, and is expanded with respect to momentum up to first order. 
The basis wavefunction based on Ref.~\cite{Zhang2013} is written as 
$\ket{\psi_{J_z=\pm1/2}}=\ket{\psi_{\pm(1/2)}^{(0)}}+\ket{\psi_{\pm(1/2)}^{(1)}}$, including the zeroth-order contribution:
\begin{equation}
\ket{\psi_{\pm1/2}^{(0)}}=u_0\ket{p_z,\uparrow(\downarrow)}+v_0\ket{p_{\pm},\downarrow(\uparrow)}
\label{eq:0order}
\end{equation}
and the first-order contribution:
\begin{eqnarray}
\ket{\psi_{\pm(1/2)}^{(1)}}=\pm k_{\pm}(iu_1\ket{p_{\mp},\uparrow(\downarrow)}+iv_1\ket{p_z,\downarrow(\uparrow)}) \nonumber\\
\mp ik_{\mp}w_1\ket{p_{\pm},\uparrow(\downarrow)}
\label{eq:1order}
\end{eqnarray}
where $u_0$, $v_0$, $u_1$, $v_1$, $w_1$ are material-dependent parameters taken to be real numbers, which control the relative ratio of different wavefunction components. The arrows denote spin angular momentum, and $(p_x,p_y)$ represent the two in-plane $p$-orbitals, with $p_{\pm}=\mp\frac{1}{\sqrt{2}}(p_x\pm ip_y)$ contributing $\pm1$ unit of angular momentum along the $z$-axis. Each term in the wavefunction separately combines orbital and spin angular momenta to give a total angular momentum $J_z = \pm 1/2$. The first-order contribution [Eq.~\ref{eq:1order}] enables the orbital character to evolve with momentum. After diagonalizing the two band Hamiltonian (Eq.~\ref{eq:kp}) and expanding under this basis wavefunction, we obtain the 2 eigenfunctions $\ket{\Phi_{1,2}}$ as a linear combination of $\ket{\psi_{\pm 1/2}}$. 
We can then derive the spin texture coupled to different $p$-orbitals to calculate the quantitative momentum evolution of spin and orbital textures. The key distinction of our methodology from the formalism of Ref.~\cite{Zhang2013} is the addition of the $\gamma$ term to the Hamiltonian in Eq.~\ref{eq:kp} which is necessary to describe the observed out-of-plane spin polarization. 
Moreover, Ref.~\cite{Zhang2013} used first principles calculations to obtain the material-specific parameters for Bi$_2$Se$_3$; such an approach is not feasible here since the electronic structure of MBT is not well-described by DFT \cite{Li2019,jiaheng2019,Zhang2019,Shikin2022}.

To overcome this challenge, we use our exhaustive dataset-- comprising the $k$-dependent orbital-resolved spin polarizations as well as the band dispersions -- to experimentally constrain the material-dependent parameters.
Our formalism, detailed in Appendix~\ref{sec:spintexture_formalism}, is based on $s$-type final states for electrons photoemitted from the TSS with 6~eV photons. The nature of the final states has been well-established for Bi-based topological insulators in the literature \cite{Ryoo2016,Bentmann2021}, including an extensive theoretical and experimental study employing spin-resolved ARPES, circular dichroism ARPES and one-step-model photoemission calculations \cite{Sanchez-Barriga2014}.
From our analysis, we obtain Hamiltonian parameters $\alpha$ and $\gamma$, and the wavefunction parameters (see Eq.~\ref{eq:0order},~\ref{eq:1order}). We show a comparison with the data in Fig.~\ref{fig:2}(k-n), where the results from the model are shown in solid lines. Table~\ref{tab:Fitting parameters} reports the parameters we obtain for all compounds. 
Next, we explain how the diverse observed behaviors are successfully reproduced by this single unified model.   

\begin{table}[h]
\caption{Parameters used for the model describing different surfaces. The units for the Hamiltonian parameters are $\alpha$ [eV$\cdot$\angstrom], $\gamma$ [eV$\cdot$\angstrom$^3$]. The wavefunction parameters $u_0$ and $v_0$ are unitless, with $u_0 \equiv 1$ to facilitate comparison of other parameter values. The higher-order wavefunction parameters $(u_1,v_1,w_1)$ have units [\angstrom]. Note that $\gamma$ has the opposite sign for MnBi$_2$Te$_4$ due to its different sample orientation (See Appendix~\ref{sec:Sz}). }
\label{tab:Fitting parameters}
\begin{ruledtabular}
\begin{tabular}{cccccccc}
 Surface &  $\alpha$ & $\gamma$ &$u_0$&$v_0$&$u_1$&$v_1$&$w_1$\\ \hline
 MnBi$_2$Te$_4$&$-1.7$& $-39$&$1$ &$1.12$&$-4.8$ &$-1.8$&$-8$\\
QL MnBi$_4$Te$_7$ & $-1.9$ & $71$ & $1$ & $0.74$ & $4.6$&$0.56$& $-16$\\
Bi$_2$Te$_3$ & $-1.9$&$69$ & $1$ & $0.06$ & $4.6$
 & $-2.4$ & $-20$\\
\end{tabular}
\end{ruledtabular}
\end{table}

Focusing on our first observation, we note that the light polarization-dependent in-plane spin $S_y$ [Fig.~\ref{fig:2}(k.l)] is consistent with previous studies on  Bi$_2$Se$_3$ \cite{Jozwiak2013,Xie2014,Zhu2014}, and has been well-understood from the zeroth-order wavefunction basis alone [Eq.~\ref{eq:0order}] \cite{Zhang2010,Zhang2013,Zhu2013,Pertsova2014,Ryoo2016}. The observation of similar $S_y$ reversal behavior in MBT further showcases the applicability of the $p$-orbital basis set.
Now we turn our attention to the second observation, which was that $S_z$ reverses direction between $p$- and $s$-pol in MnBi$_2$Te$_4$ [Fig.~\ref{fig:2}(m)], while in QL MnBi$_4$Te$_7$ [Fig.~\ref{fig:2}(n)] it does not. Such distinct $S_z$ behaviors can be quantitatively described by the basis set: the ratio of different wavefunction components (specifically $k u_1/v_0$, see Eq.~\ref{eq:0order} and Eq.~\ref{eq:1order}) controls the sign of out-of-plane spin that develops at larger $k$ for in-plane orbitals.
We also perform this analysis for Bi$_2$Te$_3$ (see Appendix~\ref{sec:BT_comp}), which has the same termination as QL MnBi$_4$Te$_7$. The similar absence of $S_z$ reversal shows that this behavior manifests the distinct wavefunction properties of each surface termination.
Furthermore, as mentioned in our third observation, the reduced $|S_z|$ when measured with $s$-pol, which is particularly pronounced for MnBi$_2$Te$_4$ [Fig.~\ref{fig:2}(m)] can also be captured by the basis wavefunction parameters (specifically the ratio $w_1/u_1$ in Eq.~\ref{eq:1order}) which tune the magnitude of $S_z$ coupled to in-plane orbitals at higher $k$.

To further confirm the validity of our model, we have also benchmarked our analysis with circular dichroism (CD) ARPES (see Appendix~\ref{sec:CD}), which has been extensively used to study conventional TIs \cite{Wang2011,Scholz2013,Cho2021} as well as MBT \cite{Yan2021,Vidal2021,Lee2023}.

\section{Spin-orbital texture of hybridized surface bands}
Next, we turn to SL terminated MnBi$_4$Te$_7$, which exhibits a more complex band structure whose assignments are even less well-understood compared to the other terminations. High-resolution ARPES shows multiple bands intersecting near $\Gamma$ \cite{Hu2020,Lee2023}, and the most prominent band, which has a parabolic dispersion, is absent in DFT \cite{Wu2020}. Our spin-resolved ARPES measurements of $S_y$ and $S_z$ measured with both $p$- and $s$-pol are shown in Fig.~\ref{fig:4}. Unlike the other studied terminations, we cannot resolve a distinct TSS due to the highly-overlapping nature of the bands and our limited $k$-resolution. Nevertheless, we show that our spin-resolved data allows us to offer insights into the observed band structure.

A leading interpretation in literature is that the outermost band belongs to a pair of spin-split bands induced by the Rashba effect, which we will refer to as ``Rashba-split" states. To study this more systematically, we plot momentum distribution curves (MDC) of spin-up and spin-down intensities at three different binding energies in Fig.~\ref{fig:4}(c), with color denoting the spin channel. The triangle markers  denote the position of not only the outermost band, but also a weaker inner band. The MDCs reveal that these two bands have parallel spin polarizations, indicating that they are not a Rashba-split pair, but states of different origins. We note that this differs from previous CD studies, where their opposite relative CD signals suggest an antiparallel alignment \cite{Vidal2021,Lee2023}. 
Nevertheless, our observation is conceptually consistent with previous interpretations which attributed the bands to hybridization between Rashba-split and Dirac states \cite{Lee2023}. This idea is illustrated in the cartoons in Fig.~\ref{fig:4}, which show the Dirac and Rashba-split states in (d) hybridizing into the states shown in (e).

Finally, we investigate the spin-orbital texture of the outer band of the Rashba-split pair \cite{Liu2016,Waugh2016,Qu2023}. 
We follow Ref.~\cite{Vajna2012} and write out the Rashba Hamiltonian $H_R$ obeying $C_{3v}$ symmetry as:

\begin{equation}
\begin{split}
H_R(k)=&E_0(k_x^2+k_y^2)+\alpha(k_x\sigma_y-k_y\sigma_x)+\\&\beta[(k_x^3+k_xk_y^2)\sigma_y-(k_x^2k_y+k_y^3)\sigma_x]+\\&\gamma(k_+^3+k_-^3)\sigma_z
\end{split}
\label{eq:Rashba}
\end{equation}

This is similar to the TSS Hamiltonian (Eq.~\ref{eq:kp}), but with an additional term with coefficient $\beta$ that has been shown to be important in systems that show prominent Rashba effects \cite{Vajna2012,Qu2023}. In addition, the $E_0$ term is spin-independent and gives a quadratic dispersion which ensures trivial topology. The data and model comparisons are shown in Fig.~\ref{fig:4}(h-i).  
Note that we exclude the range within $k=\pm0.05$~$\angstrom^{-1}$, since multiple bands overlap in this region and cannot be separated within our resolution. Remarkably, under the single band picture and same basis wavefunction set that we previously used to describe the TSSs, we can also capture the quantitative spin texture for the Rashba-split states. The parameters we use for the Rashba-split pair is separately shown in Table~\ref{tab:SL147}. 
\begin{table}
\caption{Parameters used for the model describing SL MnBi$_4$Te$_7$. The units for the Hamiltonian parameters are $E_0$ [eV$\cdot$\angstrom$^2$], $\alpha$ [eV$\cdot$\angstrom], $\beta$ and $\gamma$ [eV$\cdot$\angstrom$^3$]. The wavefunction parameters $u_0$ and $v_0$ are unitless, with $u_0 \equiv 1$ to facilitate comparison of other parameter values. The higher-order wavefunction parameters $(u_1,v_1,w_1)$ have units [\angstrom]. }
\label{tab:SL147}
\begin{ruledtabular}
\begin{tabular}{cccccccccc}
 Surface & $E_0$ & $\alpha$ &$\beta$& $\gamma$ &$u_0$&$v_0$&$u_1$&$v_1$&$w_1$\\ \hline
SL MnBi$_4$Te$_7$&$16$&$1.1$&$-8$&$33$&$1$ &$-0.56$&$1.2$ &$-1.2$&$1.9$\\
\end{tabular}
\end{ruledtabular}
\end{table}

\section{Discussion and Conclusions}

We now discuss the implications of our results for understanding the surface band structures of the MBT family. We have shown that the complex spin-orbital texture of the TSS in both MnBi$_2$Te$_4$ and QL MnBi$_4$Te$_7$ is well-described by a single-band model composed of $p$-orbitals with $J_z = \pm 1/2$. In other words, the wavefunction properties of the TSS are qualitatively similar to those of conventional TIs such as Bi$_2$Se$_3$. We emphasize the important role of spin information for arriving at this conclusion,  since the curvature of the band dispersion and reversal-behavior in CD naturally lead to more complex interpretations based on interband hybridizations \cite{Yan2021,Liang2022}. SL terminated MnBi$_4$Te$_7$, in contrast, does exhibit interband hybridizations, with the band assignments and wavefunction characters clearly demonstrated by our analysis. 
However, despite the qualitative similarity to conventional TIs, Table~\ref{tab:Fitting parameters} shows a strong material-dependence to the orbital compositions, thus pointing out the importance of systematic studies of the TSS wavefunctions. In addition, the ability to extract Hamiltonian and wavefunction parameters allows us to experimentally quantify quantum geometry related properties including Berry curvature and quantum metric \cite{Di2010,Yang2014}, which give rise to exotic topological transport behavior such as higher-order quantum Hall effects \cite{Das2023,Zhang2023}. 
 A specific case calculation is shown in Appendix~\ref{sec:BC}.
 
Beyond the interpretation of dispersions and quantum geometry, the most important question is how the wavefunction information provides insight on the issue of the gapless Dirac cone in MBT \cite{Hao2019,Chen2019,Hu2020,Nevola2020,Swatek2020, Vidal2021,Wu2020}. 
There have so far been many studies exploring the issue in literature, and the possible explanations mainly include magnetic reconstruction \cite{Hao2019,jiaheng2019,PhysRevLett.124.136407,Garrity2021,Yang2022} and defect-induced TSS relocalization \cite{Shikin2021,Tan2023,Liu2022}.
Both pictures involve factors extrinsic to the nominal magnetic and structural configuration of MBT, and they are challenging to be experimentally verified in ARPES studies.

Our wavefunction analysis suggests an intrinsic mechanism that reduces the TSS gap size compared to that predicted by DFT. When TRS is broken, the gap $\Delta$ is proportional to the strength of the Zeeman interaction between the effective magnetic field $B_{\textrm{eff}}$ generated by the Mn layer and the magnetic moment $\mu$ of the TSS \cite{Xiao2022}:
\begin{equation}
\Delta \propto \mu\cdot B_{\textrm{eff}}\propto \bra{\psi}\textbf{L}+2\textbf{S}\ket{\psi} \cdot B_{eff}
\end{equation}

\noindent where the magnetic moment includes both orbital $\textbf{L}$ and spin $\textbf{S}$ angular momenta. Since we are interested in the behavior near the $\Gamma$ point, we take the $k\rightarrow0$ limit of Eq.~\ref{eq:0order}:
\begin{equation}
\Delta\propto \frac{\abs{u_0}^2}{\abs{u_0}^2+\abs{v_0}^2}=\frac{r}{1+r}
\label{eq:u0r}
\end{equation}
where $r=u_0^2/v_0^2$ is the relative weight of out-of-plane and in-plane $p$-orbitals.

Equation~\ref{eq:u0r} reveals an  antagonistic effect of in-plane $p$-orbitals: a larger proportion of in-plane orbitals results in a smaller TSS gap. 
Based on this mechanism, we perform a direct comparison between our model and DFT calculations, focusing on MnBi$_2$Te$_4$ since in both terminations of MnBi$_4$Te$_7$ the Dirac point is hybridized with bulk bands, so the gapping behavior and orbital characters are less clearly defined.

For MnBi$_2$Te$_4$ we obtained $r=0.8$ (Table~\ref{tab:Fitting parameters}), showing a dominance of in-plane $p$-orbitals. In contrast, DFT calculations produce a value $r = 2.7$ (see Appendix~\ref{sec:DFT} for the DFT electronic structure), which shows a dominance of out-of-plane $p$-orbitals instead. Based on Eq.~\ref{eq:u0r}, this discrepancy corresponds to a $40\%$ reduction in TSS gap size compared to DFT prediction. 
In other words, DFT underestimates the in-plane orbital contribution and therefore overestimates the magnitude of the TSS gap. We emphasize that this behavior is intrinsic to the wavefunction, and it must necessarily coexist with extrinsic mechanisms (such as defects) which may additionally contribute to the gap reduction.

Finally, we briefly comment on ferromagnetic TIs, which are proposed to be more robust against disorder compared to antiferromagnetic TIs \cite{Tan2023}. Indeed, magnetic gaps have been experimentally observed in ferromagnetic bulk MnBi$_8$Te$_{13}$ \cite{Lu2021} and monolayer MnBi$_2$Te$_4$/Bi$_2$Te$_3$ \cite{Kagerer2023}, yet their magnitudes remain a factor of two smaller than theoretical predictions \cite{Otrokov2017}. This suggests that the intrinsic gap-reducing mechanism revealed in this work may be operative in ferromagnetic TIs as well, which will be the subject of future investigations.

In conclusion, our work provides a systematic study of the  MnBi$_2$Te$_4$ and MnBi$_4$Te$_7$ surface state wavefunctions. The data showcases the nontrivial orbital-coupled spin texture, and our analysis establishes the quantitative applicability of the $k\cdot p$ model for unraveling the surface electronic structure.  
Most importantly, we propose a distinct intrinsic mechanism in which the contribution of in-plane $p$-orbitals reduces the TSS gap. 
This provides crucial insights into the contradiction between theory and experiment and guides future efforts to discover and optimize magnetic topological insulators. 
\\

\begin{acknowledgements}
We acknowledge helpful discussions with D.-H. Lee, T.P. Devereaux, M. Schüler, and B. Moritz.
    This work was primarily supported by the U.S. Department of Energy, Office of Basic Energy Sciences, Division of Materials Science and Engineering under contract DE-AC02-76SF00515. Z.C.T. and Y.F.G. provided single crystals. Y. F. Guo acknowledges the National Key R$\&$D Program of China (Grant No. 
2023YFA1406100) and the Double First-Class Initiative Fund of ShanghaiTech 
University. H.X.T. and B.H.Y. performed first-principles calculations in support of this work.
\end{acknowledgements}
 
\appendix

\section{Photoemission methods}\label{sec:photomethods}

Our spin-resolved ARPES setup is based on a spectrometer with high efficiency attributed to its combination of exchange scattering for spin discrimination and time-of-flight analysis for resolving energy. 
We map out the momentum space along $k_x$ by rotating the sample with respect to axis $\theta$ (see Fig.~\ref{fig:1}(a)), with $\Gamma-K$ and $\Gamma-M$ trajectories selected by azimuthal rotation of the sample about its normal. Spin discrimination is achieved via low-energy exchange scattering from a ferromagnetic thin film (10 ML Co/W(110)), with the scattered electrons detected by a multichannel plate detector. Spin contrast is obtained by reversing the magnetization direction of the film using Helmholtz coils. The spin detector can measure the horizontal/vertical spin components (in the lab frame) by rotation of the magnetic scattering crystal, corresponding to $S_y$/$S_z$. A more detailed description of the system can be found in Ref.~\cite{Han2023}, with additional technical information in Ref.~\cite{Jozwiak2010,Gotlieb2013}. 

The laser source for spin-resolved photoemission is a commercial ytterbium fiber laser amplifier (Coherent Monaco) providing up to 80~µJ and 60~W at 1035~nm wavelength with < 300~fs pulse duration. The fifth-harmonic of the fiber laser system generates the $6$~eV photons for photoemission. \cite{Ishida2016}.

CD ARPES measurements are performed with a Scienta R4000 electron analyzer and a Ti:sapphire oscillator with photon energy quadrupling through two stages of second harmonic generation, outputting $6$~eV photons for photoemission.

\section{Sample synthesis and preparation}\label{sec:synthesis}
The MnBi$_{2n}$Te$_{3n+1}$ ($n$ = 1 and 2) single crystals in this study were grown from Mn 
(99.95$\%$), Bi (99.999$\%$), and Te (99.999$\%$) blocks by using the self-flux method. The 
mixture with a specific molar ratio of 1:8:13 was put into an alumina crucible, which 
was then sealed into a quartz tube in vacuum. The assembly was heated in a furnace 
up to $700^{\circ}$C within 10~h and kept at that temperature for 10~h, then cooled down to 
$610^{\circ}$C in 10~h and slowly further cooled down to $600^{\circ}$C and $590^{\circ}$C for $n$ = 1 and 2, 
respectively. Finally, the assemblies were taken from the furnace and immediately put 
into a centrifuge to separate black crystals with shining surfaces from the flux.

\section{Light polarization coupled spin texture formalism}\label{sec:spintexture_formalism}
Since we only focus on one of the two bands of the model, we denote the eigenfunction as $\ket{\Phi}$. We project the eigenfunction $\ket{\Phi}$ onto the desired basis to obtain the $p$-orbital coupled spin texture. We define the projections as:

\begin{equation}
c_{p_i,\uparrow(\downarrow)_j} =\bra{p_i,\uparrow(\downarrow)_j}\ket{\Phi}
\label{eq:c}
\end{equation}

\noindent Here $i$, $j$ = ${x,y,z}$ represent the $p$-orbital and spin direction. For example, $ c_{p_x,\uparrow_x}$ denotes the $\ket{p_x,\uparrow_x}$ component. The weight of the orbital $p_i$ can be then defined as:
\begin{equation}
w_{p_i}=|c_{p_i,\uparrow}|^2+|c_{p_i,\downarrow}|^2
\label{eq:w}
\end{equation}

To connect the output of this model to our measurements, we must account for the experimental geometry. With $s$-type final states (see main text), the measured spin polarization $\langle S_j\rangle$ along quantization axis $j$ can be derived as \cite{Moser2023}: 

\begin{widetext}
\begin{equation}
    \langle S_j \rangle _{p\text{-}\mathrm{pol.}}=\frac{\cos^2(\eta)(|c_{p_x,\uparrow_j}|^2-|c_{p_x,\downarrow_j}|^2)+\sin^2(\eta)(|c_{p_z,\uparrow_j}|^2-|c_{p_z,\downarrow_j}|^2)+2\cos(\eta)\sin(\eta)(Re[c_{p_z,\uparrow_j}^*c_{p_z,\uparrow_j}]-Re[c_{p_z,\downarrow_j}^*c_{p_z,\downarrow_j}]) }{\cos^2(\eta)w_{p_x}+\sin^2(\eta)w_{p_z}+2\cos(\eta)\sin(\eta)(Re[c_{p_z,\uparrow_j}^*c_{p_z,\uparrow_j}]+Re[c_{p_z,\downarrow_j}^*c_{p_z,\downarrow_j}])}      
    \label{eq:ppol}
\end{equation}
\begin{equation}
        \langle S_j\rangle_{s\text{-}\mathrm{pol.}}=\frac{|c_{p_y,\uparrow_j}|^2-|c_{p_y,\downarrow_j}|^2}{w_{p_y}}
        \label{eq:spol}
\end{equation}
\end{widetext}
\noindent where $\eta=45^{\circ}-\theta$ is the angle between the incoming laser and the sample surface normal.
The spin polarization we experimentally measure has contributions from different $p$-orbitals depending on the light polarization and sample orientation.
As can be seen from Eq.~\ref{eq:ppol}, the relative weight of $p$-orbitals probed by $p$-pol. depends on the sample rotation $\eta$. 
When $\eta=0^{\circ}$, spin polarization probed by $p$-polarized light is solely due to the $p_x$ orbital contributions. When $\eta=90^{\circ}$, only $p_z$ orbitals contribute to the signal. On the other hand, according to Eq.~\ref{eq:spol}, $s$-polarized light solely probes $p_y$ orbital contributions, and is independent of the angle between sample normal and incident light. After computing the momentum-dependent spin-polarization from this model, the final result is convolved with a Gaussian function to take into account the finite momentum resolution in the experiment. 

To highlight the need for these corrections related to experimental geometry, we note that Eq.~\ref{eq:ppol} shows that the spin polarization measured with $p$-pol. is asymmetric with respect to $k=0~\angstrom^{-1}$ due to the way contributions from $p_x$ and $p_z$ orbitals are mixed as the sample is rotated. In contrast, $s$-pol. data should be symmetric in magnitude with respect to $k=0~\angstrom^{-1}$.  These behaviors are indeed observed in the experimental data and well-described by our model (Figs.~\ref{fig:2} and \ref{fig:4}).

\section{Connection to Circular Dichroism ARPES}
\label{sec:CD}
\begin{figure}
    \includegraphics{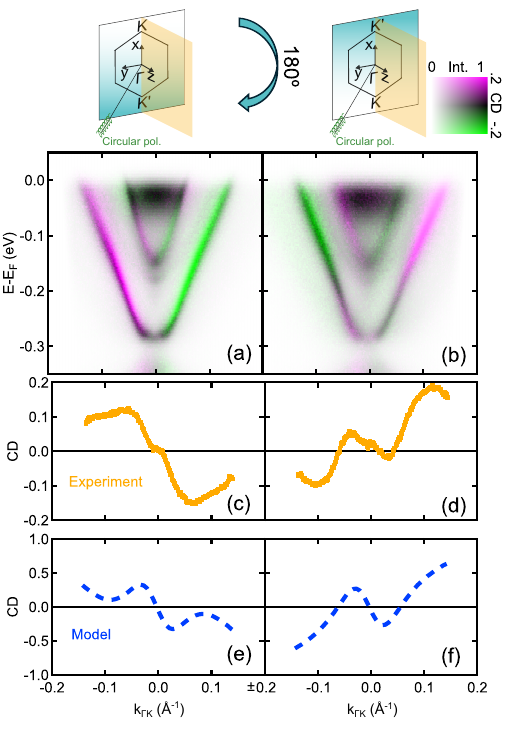}
    \caption{Measurement geometry for circular dichroism (CD)-ARPES of QL MnBi$_4$Te$_7$. Circularly-polarized light is incident at $50^{\circ}$ with respect to the surface normal. $\Gamma-K$ is along the $x$-axis.  The yellow plane indicates the plane from which the electrons are collected. (a) and (b) show the CD-ARPES data before and after $180^{\circ}$ rotation of the sample, plotted in the 2D colorscale as shown.  (c,d) CD of the surface state before and after sample rotation. (e,f) CD yielded from our model with wavefunction parameters derived from the spin-resolved measurements. }
    \label{fig:3}
\end{figure}
\begin{figure}
\includegraphics{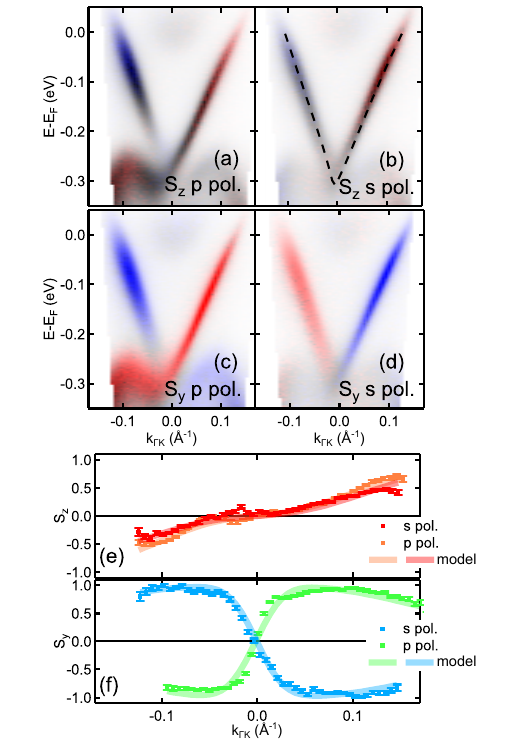}
\caption{Data for Bi$_2$Te$_3$ taken at room temperature along $\Gamma-K$. Specific measurements are $S_z$ with (a) $p$-pol and (b) $s$-pol, and $S_y$ with (c) $p$- pol and (d) $s$-pol. (e) $S_z$ with both $p$ and $s$- pol from spin image (dots), overlapped with result yielded from model(lines). (f) Same but for $S_y$.}
 \label{fig:BT}
\end{figure}
\begin{figure*}
 \includegraphics{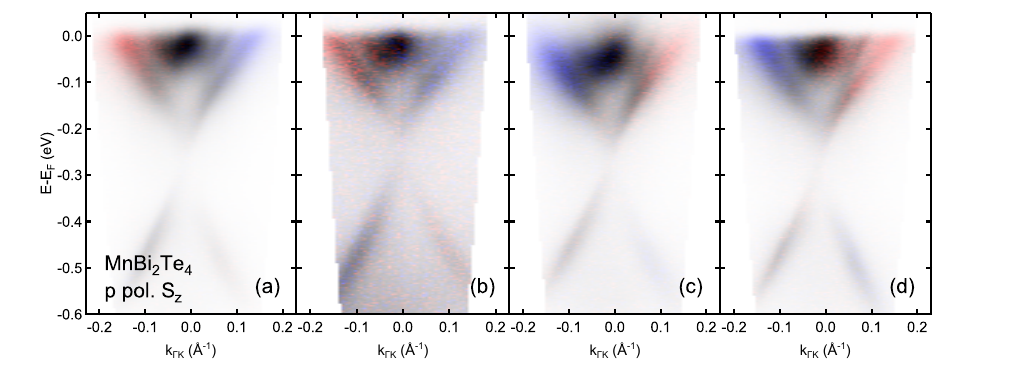}
 \caption{Four different measurements of MnBi$_2$Te$_4$. Out-of-plane spin polarization $S_z$ taken with $p$-polarized light at temperature (a) 80~K, (b) 80~K, (c) room temperature, and (d) 18~K. Note that $S_z$ in (a,b) is opposite to that in (c,d) because of a $180^{\circ}$ difference in the azimuthal sample orientation, which produces a reversal of the out-of-plane spin components (see Fig.~\ref{fig:1}(b)).}
 \label{fig:all124}
\end{figure*}
\begin{figure}
 \includegraphics{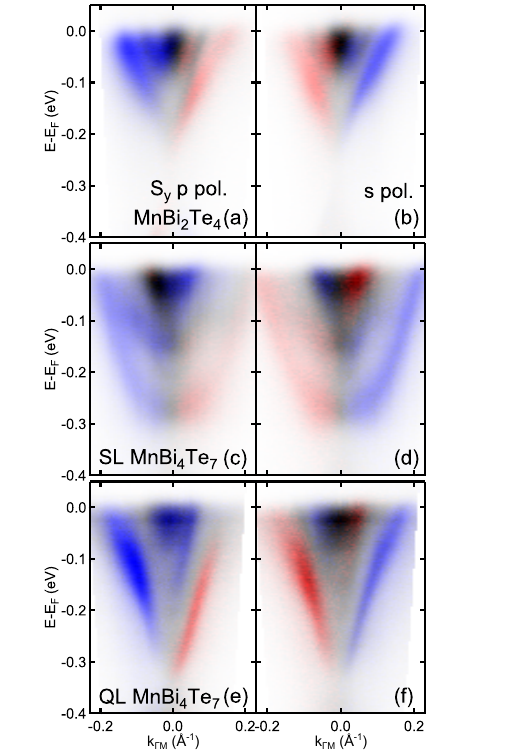}
 \caption{(a-b) $S_y$ along $\Gamma$M for MnBi$_2$Te$_4$, taken with $p$ and $s$ polarized light. (c-d) Same but for SL MnBi$_4$Te$_7$ and (e-f) QL MnBi$_4$Te$_7$. Colorscale same as main text figures.}
 \label{fig:GMIP}
\end{figure}

To show that the spin-orbital texture of the TSS wavefunction is correctly described by the model in the main text, we here benchmark the model by using it to compute a separate experimental observable: circular dichroism (CD). 
We perform CD-ARPES on QL MnBi$_4$Te$_7$ using the measurement geometry shown in Fig.~\ref{fig:3}, where there is a $50^{\circ}$ angle between the incident circularly polarized light and sample normal (see Appendix~\ref{sec:photomethods}). We repeat the measurement after rotating the sample azimuthally by $180^{\circ}$ to test the consequences of the $C_{3v}$ symmetry. Fig.~\ref{fig:3}(c,d) presents the CD signal along the TSS, which depends on sample orientation.

To calculate CD, we assume circularly polarized light projects the wavefunction onto the orbital angular momentum basis $\ket{p_{\pm}}$, which refers to $m_l = \pm 1$ along the axis defined by the light propagation direction \cite{Park2012}. Based on the geometry shown in Fig.~\ref{fig:3}, this basis can be written: 
\begin{eqnarray}
\ket{p_{\pm,\uparrow(\downarrow)}}=&\mp\frac{1}{\sqrt{2}}[\ket{p_{x,\uparrow(\downarrow)}} 
\pm \nonumber\\ &i(\cos(\kappa)\ket{p_{y,\uparrow(\downarrow)}}+\sin(\kappa)\ket{p_{z,\uparrow(\downarrow)}})]
\end{eqnarray}

\noindent where $\kappa=50^{\circ}$ is the angle between the light propagation and the sample normal. As a result, the total projection on the orbitals for plus and minus circularly polarized light can be written as 
\begin{equation}
\mathrm{CD}_{\pm}=|\bra{p_{\pm,\uparrow}}\ket{\Phi}|^2+|\bra{p_{\pm,\downarrow}}\ket{\Phi}|^2
\end{equation}
Hence, the normalized circular dichroism  is given as:
\begin{equation}
\mathrm{CD}=\frac{\mathrm{CD}_+-\mathrm{CD}_-}{\mathrm{CD}_++\mathrm{CD}_-}
\end{equation}

The calculation results, plotted in Fig.~\ref{fig:3}(e,f), qualitatively reproduce the CD observed experimentally, including the azimuthal angle-dependence. 

Intuitively, CD measures the OAM projected onto the axis of light propagation, which in this geometry is a linear combination of $y$- and $z$- directions. The behavior at higher $k$ is associated with the OAM derived from the $\gamma$-term of the Hamiltonian, which obeys threefold rotational symmetry and therefore reverses sign upon 180$^\circ$ rotation \cite{Wang2011,Ketterl2018}. The composite effects of in-plane and out-of-plane OAM lead to a complex CD texture that depends sensitively on the light-sample geometry. The fact that this is captured by our wavefunction evidences the intricate coupling of orbital and spin angular momenta, and further corroborates the validity of our model.

\section{Comparison to Bi$_2$Te$_3$}\label{sec:BT_comp}

We perform similar wavefunction analysis for Bi$_2$Te$_3$, the results of which are shown in Fig.~\ref{fig:BT}. The model again captures the measured spin texture. It can be seen that $S_y$ flips [Fig.~\ref{fig:BT}(f)], while $S_z$ does not flip between $p$-pol. and $s$-pol. [Fig.~\ref{fig:BT}(e)]. Such behavior is the same as QL terminated MnBi$_4$Te$_7$ as discussed in the main text [Fig.~\ref{fig:2}(i,j)], which further showcases the distinct wavefunction properties of each surface termination. See Table~\ref{tab:Fitting parameters} for the model parameters for Bi$_2$Te$_3$.

\section{Out-of-plane spin texture along $\Gamma-K$}\label{sec:Sz}

Figure.~\ref{fig:all124} shows different sets of $S_z$ data of MnBi$_2$Te$_4$ along $\Gamma-K$ taken on different cleaves at different times and temperatures. Each time it was prepared and remounted so the orientation could be integer multiples of $60^{\circ}$ off from each other, due to the hexagonal symmetry of the Fermi surface. It can be seen that $S_z$ for Fig.~\ref{fig:all124}(a,b) is opposite to  Fig.~\ref{fig:all124}(c,d). The difference in $S_z$ depends on the orientation of the sample.

\section{In-plane spin texture along $\Gamma-M$}\label{sec:GM}

We show in Fig.~\ref{fig:GMIP} the tangential in-plane spin measurements $S_y$ of MnBi$_2$Te$_4$ and MnBi$_4$Te$_7$ along $\Gamma-M$. We rotate the sample azimuthally so that $\Gamma-M$ is along $k_X$ as shown in Fig.~\ref{fig:1}(a). Under the constraint of time reversal symmetry and mirror symmetry along $\Gamma-M$, there is no $S_z$ allowed in the paramagnetic phase, therefore here we only focus on $S_y$. The measured $S_y$ along $\Gamma-M$ [Fig.~\ref{fig:GMIP}] agrees with that measured along $\Gamma-K$ qualitatively, revealing the helical spin texture of the surface states.

\section{Extracting the Quantum geometry: Berry Curvature and Berry curvature multiples}\label{sec:BC}
\begin{figure}
\includegraphics{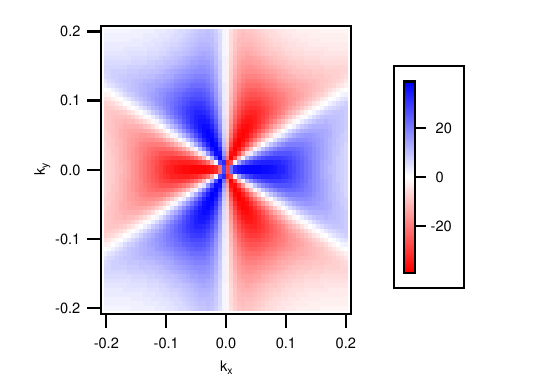}
\caption{Simulated Berry curvature of QL MnBi$_4$Te$_7$.}
\label{fig:BC}
\end{figure}

The extracted wavefunction parameters can be used to quantify quantum geometry, which is the basis for the quantum anomalous and nonlinear Hall effects.
Here we provide a simple example 
by calculating the Berry curvature. Based on the Hamiltonian Eq.\ref{eq:kp} and Ref.~\cite{Li2014}, the Berry curvature $\Omega$ can be calculated as:
\begin{equation}
\Omega (k_x, k_y)=\frac{\alpha^2(k_x(-3k_y^2+k_x^2)\gamma)}{(\alpha^2(k_x^2+k_y^2)+k_x(k_x^2-3k_y^2)\gamma)^{1.5}}
\label{eq:BC}
\end{equation}
where $\alpha$ and $\gamma$ are the Hamiltonian parameters extracted from our analysis. 

Fig.~\ref{fig:BC} shows the calculated $\Omega$ for QL MnBi$_4$Te$_7$. In addition, Berry curvature multipoles (dipole, quadrupole, hexapole etc) ~\cite{Zhang2023} and quantum metric (the real part of quantum geometry) can also be derived from a similar manner \cite{Das2023}. 

While local Berry curvature in Fig.~\ref{fig:BC} is finite, the integral across momentum space is zero due to time reversal symmetry. However, the integral of higher order multipoles can still be nonzero, and hence can lead to nonlinear Hall conductivities \cite{Sodemann2015}.
Therefore, our methodology provides a quantitative pathway to extract nontrivial phenomena related to quantum geometry.

\section{DFT calculation}\label{sec:DFT}
\begin{figure}
\includegraphics{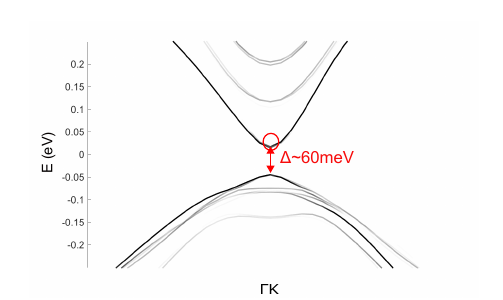}
\caption{DFT slab calculations of the band structure of MnBi$_2$Te$_4$. The red circle denotes the $\Gamma$ point from which the $p$-orbital weights are extracted. The red arrow represents the TSS gap, with the gap size $\Delta$ also labeled.}
\label{fig:DFT}
\end{figure}

DFT calculation on the surface states of MnBi$_2$Te$_4$ is performed using slab models. The MnBi$_2$Te$_4$ slab comprises five septuple layers. To prevent interactions between periodic images, a vacuum thickness of at least 12~$\angstrom$ was included in all slab models. We considered intra-layer ferromagnetic and inter-layer antiferromagnetic configurations, with local magnetic moments collinearly aligned along the out-of-plane direction.
All calculations were performed using density functional theory (DFT) as implemented in the Vienna Ab-initio Simulation Package (VASP) \cite{Kresse1996}. The Perdew-Burke-Ernzerhof (PBE) generalized gradient approximation \cite{Perdew1996} was employed for the exchange–correlation functional. A plane-wave energy cutoff of 350~eV was used, and the two-dimensional Brillouin zone was sampled with a k-point grid denser than $6\times 6$. A Hubbard U (5~eV) correction was applied to the Mn 3d orbitals. Spin–orbit coupling was included in all electronic structure calculations to ensure accurate description of the surface state properties.

Fig.~\ref{fig:DFT} shows the electronic structure of MnBi$_2$Te$_4$ near $\Gamma$ along $\Gamma-K$. The intensity represents the weight projected to the surface. We directly extract the $p$-orbital weights of the TSS at $\Gamma$, at the energy-momentum location indicated by the red circle.

\newpage
%

\end{document}